\documentclass{article}        
\usepackage{epsfig,bm,float}

\usepackage{wrapfig}
\usepackage[table,xcdraw]{xcolor}
\usepackage{graphicx}
 \usepackage{setspace}
\usepackage{amsmath}
\usepackage{amssymb} 
\usepackage{physics}
\usepackage{braket}
\usepackage{fullpage}
\usepackage{setspace}
\begin{document}               

\begin{center}
{\setstretch{1.1}
\noindent {\Large \textbf{A Method for Preparation and Readout of Polyatomic Molecules in Single Quantum States}} \\
\vspace{0.1in}
}
{David Patterson \\ davepatterson@ucsb.edu}
\end{center}

Polyatomic molecular ions contain many desirable attributes of a useful quantum system, including rich internal degrees of freedom and highly controllable coupling to the environment. To date, the vast majority of state-specific experimental work on molecular ions has concentrated on diatomic species. The ability to prepare and readout polyatomic molecules in single quantum states would enable diverse experimental avenues not available with diatomics, including new applications in precision measurement, sensitive chemical and chiral analysis at the single molecule level, and precise studies of Hz-level molecular tunneling dynamics. While cooling the motional state of a polyatomic ion via sympathetic cooling with a laser cooled atomic ion is straightforward, coupling this motional state to the internal state of the molecule has proven challenging.  Here we propose a new method for readout and projective measurement of the internal  state of a trapped polyatomic ion.  The method exploits the rich manifold of technically accessible rotational states in the molecule to realize robust state-preparation and readout with far less stringent engineering than quantum logic methods recently demonstrated on diatomic molecules. The method can be applied to any reasonably small ($\lesssim$ 10 atoms) polyatomic ion with an anisotropic polarizability. 
  

\section{Introduction}
The last half decade has seen enormous progress in our ability to prepare, cool, and read out the state of trapped molecular ions.  The vast majority of this work has concentrated on diatomic molecules, and primarily on hydrides\cite{jusko2014two,asmis2003gas,putter1996mass,lien2014broadband,staanum2010rotational,hechtfischer2002photodissociation,roth2005production,shi2013microwave,cirac1992laser,hume2011trapped,wester2009radiofrequency}. Neutral polyatomic molecules have been prepared in single internal states and cooled to the mK regime\cite{zeppenfeld2012sisyphus}.  Much of this work has been motivated by the potential to realize high resolution rotational and vibrational spectroscopy in such systems\cite{rugango2016vibronic}.  Diatomic molecular ions, such as CaH$^+$, are natural candidates for researchers studying the corresponding \emph{atomic} ion (e.g. Ca$^+$), but exhibit technically challenging rotational spectra with levels spaced by 100s of GHz. 
The internal state of trapped molecular ions was recently measured nondestructively, via quantum-logic spectroscopy\cite{wolf2016non}. In 2017 Chou et al. demonstrated state-selective detection and heralded projective preparation of trapped CaH$^+$ ions, using only far-off resonant light\cite{chou2016preparation}.  That work further demonstrated quantum logic assisted optical pumping of magnetic substates within a rotational manifold, opening a general pathway to a single state preparation.
Sympathetic cooling of trapped molecular ions via collisions with cold \emph{neutral} atoms is a rapidly evolving field\cite{hudson2016sympathetic}.  The short-range collisions intrinsic to this approach are both an asset and a liability - while these collisions can in principal cool all degrees of freedom of a molecular ion, they also allow for chemical reactions and potentially unwanted inelastic collisions leading to neutral atom trap loss.  

Polyatomic molecular ions have been cooled to millikelvin motional temperatures but have never been prepared in a single ro-vibrational state\cite{ostendorf2006sympathetic,wellers2011resonant,mokhberi2014sympathetic}. Ensembles of neutral polyatomic molecules have been electrically trapped, prepared in known internal states, and motionally cooled to the mK regime\cite{zeppenfeld2012sisyphus}. Remarkably, the use of state-dependent heating of trapped ion ensembles as a spectroscopy method was first suggested by Hans Dehmelt in 1968, long before the invention of laser cooling\cite{dehmelt1968bolometric}


The ability to control polyatomic molecules at the single quantum state level opens research pathways substantially beyond what diatomic molecules offer.  Parity violation is predicted to produce enantiomer-dependent shifts in the spectra of chiral molecules, but these tiny shifts have never been observed\cite{quack2008high}.  The precise spectroscopy enabled by the methods proposed here would allow direct observation of this effect, as well as coherent molecular dynamics on the few-Hz level.  Vibrational spectroscopy of homonuclear diatomic molecular ions has been proposed for high-sensitivity searches for time-variation of the electron-nucleon mass ratio, and Raman-active modes of nonpolar, closed shell triatomic molecular ions such NO$_2^+$, which could be observed directly by the system described in this work, appear to be a highly systematic-immune platform for such searches\cite{hanneke2016high}. 
Polyatomic molecules have been suggested as attractive systems in searches for both nuclear and electronic permanent electric dipole moments\cite{kozyryev2017precision}.  The single molecule sensitivity and lack of requirements of well-behaved electronic transitions make the proposed system attractive for studying exotic species where high fluxes are prohibitive, such as molecules containing the short-lived $^{225}$Ra nucleus, which are predicted to be orders of magnitude more sensitive to new physics than current searches.  Small asymmetric top molecules typically exhibit a rich spectrum of allowed transitions from $<$ 1 GHz to $>$ 100 GHz, making them attractive and flexible components in proposed hybrid schemes to couple molecules to other quantum systems, such as superconducting microwave resonators\cite{andre2006coherent}.
Finally, the methods described here provide a general and in principle portable chemical analyzer that can definitively and non-destructively identify both isomer and enantiomer of individual molecules - an analytical chemistry feat which is substantially beyond our current state of the art. 



  \begin{figure}[h!]
\includegraphics[width=6in]{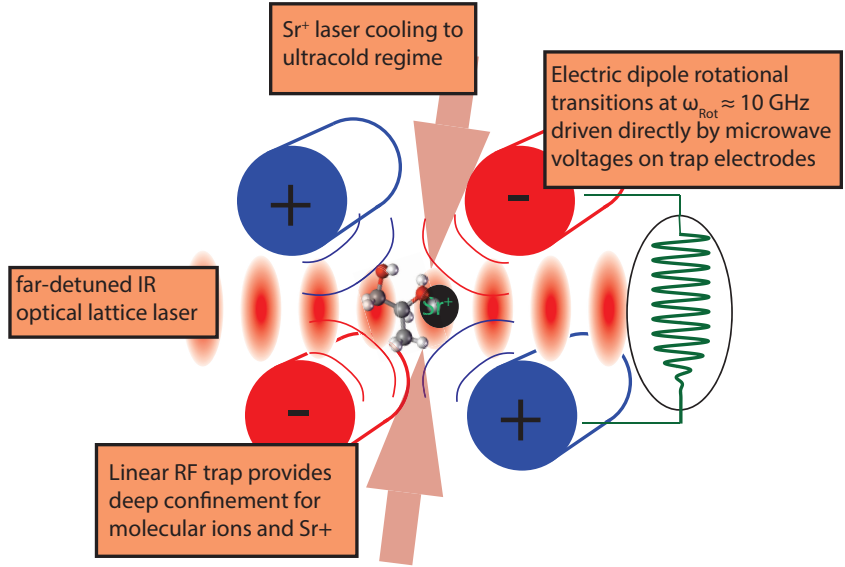}
\caption{A sketch of the proposed apparatus, not to scale.  The configuration is a linear Paul trap (axis out of the page).  A small crystal comprised of a single Sr$^{+}$ ion and a single molecular ion is aligned along the trap axis.  Two endcap electrodes (not shown) provide axial confinement. The motion of the ensemble is cooled via laser cooling applied to the Sr$^{+}$ atoms, and rotational transitions within the molecule can be driven via microwave frequency electric fields applied directly to the trap electrodes.  In addition, an optical lattice couples the molecule's internal state to its motional state.  The lattice is linearly polarized (vertical in the figure).  This coupling can be used to read out the molecule's state by heating the ensemble in a molecule-state dependent way.  Both the cooling lasers and the optical lattice propagate in a direction that is not parallel to any principal axis of the trap. Realizations in spherical Paul traps or axially laser-cooled Penning traps are also possible\cite{van2013relaxation}. 
\label{appfig}}
\end{figure}

\section{Apparatus}

The basic components of our proposed apparatus are shown in figure \ref{appfig}.  The apparatus consists of a linear Paul trap, upon which is superimposed an optical lattice, formed by two counterpropagating, linearly polarized infrared laser beams.  These lasers could be a single retro-reflected laser, which allows for maximal vibrational stability but less experimental control, or separate counter-propagating lasers brought in via separate objectives. Both the cooling lasers and the optical lattice propagate in a direction that is not parallel to any principal axis of the trap, to allow for addressing of multiple motional modes.
The Paul trap is comprised of the four rods carrying an RF voltage providing radial confinement, and two endcaps carrying a DC voltage to provide axial confinement.  A small, static magnetic field of a few Gauss defines a quantization axis, which is required for Sr$^{+}$ laser cooling; this field can be in any direction as long as it is not parallel to the propagation axis of the Sr$^{+}$ cooling lasers\cite{berkeland2002destabilization}. The complex dynamics in the hybrid ion trap-optical lattice potential have been previously studied, and is a model system for studying friction\cite{pruttivarasin2011trapped}

In addition to the trapping fields and optical potential, trapped ions can also be addressed via GHz-frequency electric fields applied directly to the trap electrodes.  
These high-frequency fields will have negligible effect on trapped atomic ions, but can rapidly (Rabi frequency $\Omega_R >$ 10 MHz) drive electric dipole transitions between rotational states of polar molecular ions.  These electric fields can be applied in arbitrary polarization by appropriate choice of electrodes.  If three electrodes (e.g. the top two in figure \ref{appfig} and one endcap) are separately addressed via high frequency electronics, arbitrary AC fields in $\hat{x}$, $\hat{y}$, and $\hat{z}$ can be applied.  This ability to agilely drive arbitrary transitions between rotational states is straightforward in molecules with 5-10 atoms, but technologically difficult in diatomics, and especially hydrides, where the required frequencies are in the challenging $\sim$ 200 GHz range.  

The entire apparatus can be cooled to $T \sim $ 4 K, and cold helium or hydrogen buffer gas can be introduced into the trap volume.

 \subsection{A state dependent potential and state dependent heating}
While the pseudopotential formed by the Paul trap depends only on the mass and charge of the trapped ions, and is therefore state-independent, the potential formed by the optical lattice is highly state dependent in molecules with anisotropic polarizabilities.   

The optical potential is given by
\[
U_{optical} = \frac{\alpha_{eff} I_0 }{ c \epsilon_0}\cos{\frac{4 \pi (z-z_0)}{\lambda}}
\] 
 where $I_0$ is the optical intensity, $\lambda$ is the wavelength, $z_0$ is the distance from the lattice maximum to the ion trap center, and $\alpha_{eff}$ is the effective polarizability. $\alpha_{eff}$ depends on the anisotropic polarizability of the molecular ion, but typically varies by order 10\% between molecular states of distinct $m_J$.  
 

The polarizability anisotropy  is characterized by a dimensionless parameter $s \equiv (\alpha_{\parallel} - \alpha_{\perp}) / \bar{\alpha}$. While a complete description of the polarizability in asymmetric top molecules requires a rank 2 tensor (which is not necessarily diagonal in the principal inertial axes of the molecule), the physics relevant for this proposal only relies on a non-zero anisotropy $s$. $s$ is expected to be order of unity for all molecules with non-spherical geometry, including all linear molecules and all planar molecules, and has been calculated for a wide range of neutral species\cite{ewig2002ab}.  The anisotropic polarizability of molecules is the mechanism responsible for rotational Raman transitions\cite{atkin2006atkins}.

While dynamics within the hybrid potential $U_{total} = U_{optical} +U_{ion}$ are complex, at low excitation energies ($E << U_{optical}$) they are characterized by a frequency $\omega_{trap}$ which is on the order of the faster of the ion trap secular frequency $\omega_{secular}$  and the optical trap secular frequency $\omega_{lattice} = \left(\frac{16 \pi^2 U_{optical}}{\lambda^2 m}\right)^{1/2}$.  The tight focus and short length scale of the optical lattice allows substantial optical forces to be applied to the molecule.   For the parameters listed in table \ref{parametertable}, a typical molecule in the lattice feels an acceleration of $|a_{optical}| \approx  10^5$ m s$^{-2}$.

 Figure \ref{classical_sim}A shows the potentials felt by a molecule in each of two states, for reasonable estimated molecular constants and trap parameters as described in table \ref{parametertable}.    Polar molecular ions can be agilely driven between these two states via microwave frequency electric fields. The rich energy spectrum of polyatomic molecules, and in particular asymmetric top molecules, is highly interconnected, meaning that any two states of energy $E \lesssim 10 k_B$ ($E/h \lesssim 200 $ GHz) can typically be connected by a series of low frequency ($\omega \lesssim 2\pi \times 20$ GHz) electric dipole transitions (figure~\ref{biggrot} 

This combination of fields - a state-independent trap, a rotational state-dependent potential superimposed on this trap, and agile microwave drive that allows for controlled transitions between rotational states - allows large, state-specific forces to be applied to trapped molecular ions.  These forces can be used to motionally heat trapped ensembles conditional on the internal state of the molecular ions.

 
   Several groups have realized state manipulation and readout of diatomic molecular ions via state-dependent potentials \cite{leibfried2012quantum,ding2012quantum,wolf2016non,staanum2010rotational}.  The work of Chou et al. is particularly relevant, as they rely only on an anisotropic polarizability in their molecular ions, rather than a specific electronic structure as is required if near-resonant optical fields are used\cite{chou2016preparation}.  Sympathetic heating spectroscopy, which is closely related to the readout method proposed here, has been demonstrated in small crystals of atomic ions \cite{clark2010detection}.  In contrast to that work, the heating proposed here does not require any absorption of optical photons by the ``dark'' species, and spontaneous emission of the dark [molecular] species plays no role. To our knowledge, the proposed hybrid trap and the method proposed here to realize robust, state-dependent heating is novel.

\section{State readout method}\subsection{Ensemble preparation}
 The molecular ion of interest is co-trapped with a laser-coolable atomic ion, such as Sr$^{+}$, in the trap shown in figure \ref{appfig}.  A small crystal, comprised of a single molecule and a single Sr$^{+}$, is considered for simplicity.  Extensions to larger crystals comprised of many molecular ions and many atomic ions are also possible, and are discussed in section \ref{bigcrystalsection}.


The ions are 
first buffer gas cooled to a temperature of about 10 K.  Cryogenic buffer gas cooling is a versatile tool for cooling internal degrees of freedom of both neutral and charged species, and appears to work more efficiently on larger (non-diatomic) species\cite{endres2017incomplete,drayna2016direct,hansen2014efficient}. Micromotion effects are known to lead to heating in ion-neutral buffered mixtures, but such effects are minimized when $m_{neutral} \ll m_{ion}$, as is expected here\cite{chen2014neutral}. Helium is the most natural buffer gas, but H$_2$ is also possibile; H$_2$ offers higher base temperatures and likely more problematic clustering behavior, but can be reliably cryopumped away to essentially perfect vacuum at temperatures $\lesssim$ 8 Kelvin.  While the primary advantage of the cryogenic environment is the cooling of the internal state of the molecules, the cold environment also provides very high vacuum and a low black body radiation temperature, resulting in long rotational state lifetimes ($\tau \gg 1$ s). 

Buffer gas cooling is necessary because molecules with rotational constants in the 2-10 GHz range, corresponding to approximately 5-15 atoms, occupy many thousands of ro-vibrational states at room temperature, making any state-specific cooling scheme infeasible.  The achievable temperature of $\sim$10 K is significantly above the ``single rotational state'' temperature $T_{rot} \approx 2B/k_B \approx $ 300 mK, but rotationally cooled molecules will be left in a small number of states ($\lesssim 50$) which can be addressed individually with microwave fields or optical fields modulated at microwave frequencies.

Once the molecules have collisionally thermalized with the buffer gas, the buffer gas is removed via a cryopump and the motion of the ensemble is cooled via laser cooling applied to the Sr$^{+}$ ion.  Absolute ground state cooling is not required to implement the methods proposed here, significantly relaxing trap design requirements.

 The laser cooled ensemble is now motionally cold  ($T_{motional} <$ 1 mK),  while the internal temperature of the molecular ion remains at $\sim$~10 K.  The molecular ion will be subject to electric fields from both the trapping fields and from the motion of the nearby atomic ion, but these fields vary slowly compared to the $\sim$ 10 GHz frequencies of molecular rotation, and are therefore extremely unlikely to change the molecule's internal rotational state.  The molecule is therefore left in one of many internal states $\ket{0}..\ket{n}$ with $E_n/k_B \lesssim 10$ K.  It is the task of our readout method to determine which one.
 
 

 \subsection{State-specific heating}
 \label{bolometrysection}
 The ideal readout method would projectively measure which state among $\ket{0}..\ket{n}$ the molecule is in. The method described in this section, which contains the essential physics of this proposal, achieves a slightly more modest goal: we will choose two states, $\ket{0}$ and $\ket{1}$, and determine if the molecule lies within the space spanned by $\ket{0}$ and $\ket{1}$, which we denote $\ket{0,1}$.  Extensions to single state determination are described in section \ref{singlestatesection}. We treat in this section the case in which states $\ket{0}$ and $\ket{1}$ are non-degenerate, but extensions to the more realistic case where $\ket{0}$ and $\ket{1}$ are degenerate or nearly degenerate are straightforward, and presented in section \ref{largesubspacesection}.
 
 Our method, which could be considered a version of the sympathetic heating spectroscopy demonstrated by Clark et al., will work by applying an effective potential to the molecules which is constant for molecules outside $\ket{0,1}$, and time varying at about the trap secular frequency for molecules within $\ket{0,1}$\cite{clark2010detection}.  In the event that the molecule began within $\ket{0,1}$, the ensemble is heated, and the molecule is left in a random internal state within $\ket{0,1}$.  In the event that the molecule began in a state $\ket{k}$ outside $\ket{0,1}$, the ensemble is not heated and the molecule is left in $\ket{k}$. 
 
 The complete sequence to determine if the molecule lies within $\ket{0,1}$ is as follows:

 \begin{enumerate}
\setlength \itemsep{-5pt}

  \item The trapped ensemble is laser cooled to a low motional temperature $T$ via the Sr$^+$ ion; ideally this would be the motional ground state, but this is not a requirement.  The optical lattice is turned off during this step.

  \item The cooling lasers are turned off, and an optical lattice is adiabatically applied to the ensemble via far-detuned infrared lasers.  
  While the potential realized by the combination of the ion trapping fields and the optical lattice is state-dependent, the net potential is conservative within the pseudopotential approximation. The dynamics within such a potential are complex, but are characterized by motion on a timescale of order $\tau \approx \omega_{trap}^{-1}$.  
  
  \item A pulsed, microwave frequency electric field $E_{mw}$ at frequency $\omega_{01}$ is applied.  Molecules outside of $\ket{0,1}$ are not affected by this field; for these molecules, the potential remains constant in time, and the molecules are not heated.  Molecules in $\ket{0}$ are driven to $\ket{1}$ and vice-versa, with a Rabi frequency $\Omega_R$ chosen to be comparable to $\omega_{trap}$ or greater.    Molecules which began in $\ket{0,1}$ are therefore moved between $\ket{0}$ and $\ket{1}$, which feel different optical potentials, and the ensemble is motionally heated. The duration of $E_{mw}$ is chosen such that an approximate $\pi$-pulse $\ket{0} \leftrightarrow \ket{1}$ is driven, but fidelity of this swap is not critical.
  
    \item Additional $\pi$-pulses are applied in a pseudorandom sequence, pushing molecules between $\ket{0}$ and $\ket{1}$ on a time scale comparable to  $\tau \approx \omega_{trap}^{-1}$.  As molecules are driven between potentials, they feel a time-varying force and are heated.  A typical pseudo-random sequence would be additional $\pi$ pulses delivered with a Poisson distribution with rate $\Gamma_{flip} \lesssim \omega_{trap}^{-1}$
    
  
  \item The microwave drive is turned off, and the optical lattice is adiabatically lowered.
  
  
  \item The ensemble motional temperature is read out via the motion of the Sr$^+$ ion. This temperature now depends strongly on whether the molecule lies within $\ket{0,1}$. Temperature readout will not change the internal state of the molecular ion.
  
  \item Steps 1-6 can be repeated to increase fidelity.
  \end{enumerate}
  
\begin{figure}[h!]
\includegraphics[width=6in]{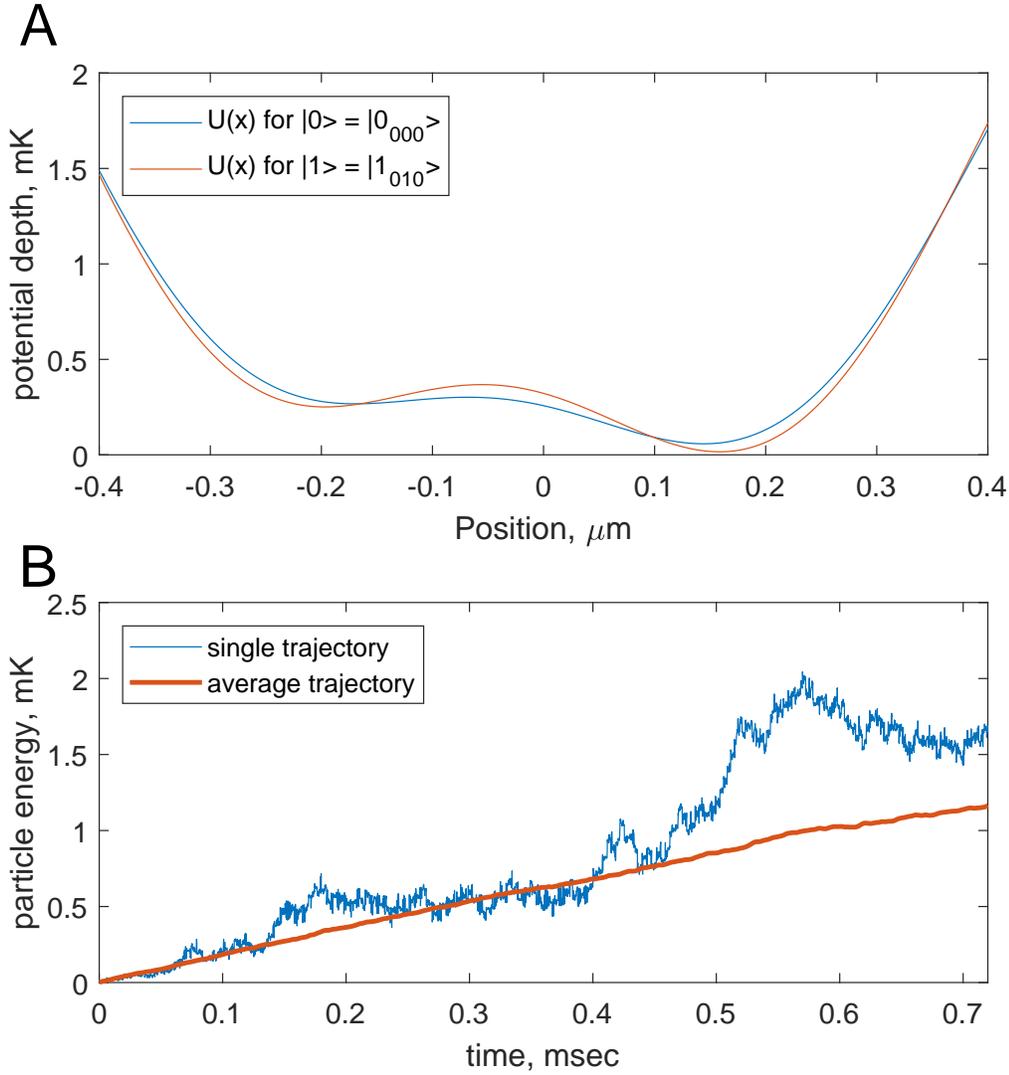}
\caption{A: The state-dependent potential seen by a polyatomic molecule from the combined Hamiltonian $H_c=H_{trap} + H_{optical}$, for parameters listed in table \ref{parametertable}.  Asymmetric top rotational states are labeled via $\ket{J_{k_ak_cm_J}}$. This highly tunable, non-separable Hamiltonian allows for state-specific, motional sideband addressing of a wide variety of polyatomic molecules.  B: The blue trace shows a classically simulated trajectory of a one atom, one-molecule ensemble trapped in three dimensions driven randomly between the potentials shown in A.  The red trace shows the average of many trajectories, reflecting a heating rate of $\sim$ 1.5 K s$^{-1}$. 
\label{classical_sim}} 
\end{figure}

  The heating described in steps 3 and 4 can be understood in the following semi-classical picture.  Consider a motionally cold molecule initially in $\ket{0}$ (blue potential in figure \ref{classical_sim}A).  A $\pi$ pulse is applied, putting the molecule in $\ket{1}$ (red potential in figure \ref{classical_sim}).  The ensemble is motionally excited in this new potential, and will fall ``downhill'' towards the new trap minimum.  
  
  The energy delivered can be increased by applying additional microwave pulses.  Each time a $\pi$ pulses is applied, the molecule will again switch potentials.  The microwave drive is intentionally dithered to ensure that subsequent kicks are uncorrelated, although in practice complex trap dynamics will likely ensure incoherent behavior even without explicit randomization. Although the details of the dynamics are complex and likely in practice  uncontrollable\cite{pruttivarasin2011trapped}, molecules which began this process in $\ket{0}$ or $\ket{1}$ will receive a sequence of effectively random momentum kicks, heating the ensemble; molecules which began in a different state $\ket{k}$ will not be addressed by the microwave fields and will not be heated.    
  This selective heating method reduces to a variant of QLS in the simultaneous limits of resolved sidebands, ground state cooling, a single, well-controlled $\pi$-pulse drive, and high fidelity single phonon readout, but our method  provides high-fidelity readout under much less stringent conditions than QLS.  In particular, neither sideband resolution nor motional ground state cooling is required. The proposed method does not rely on the applied optical potential $U_{optical}$ being a small perturbation on the trap pseudopotential; for the parameters listed in table~\ref{parametertable} and figure \ref{classical_sim}A it is not.   
  
  The above semi-classical description is of course incomplete; in fact, many motional states and the internal states $\ket{0}$ and $\ket{1}$ will be thoroughly mixed by the series of pulses, and molecules will in general not be in a pure internal state.  The description further ignores the spatially-dependent detuning of the $\ket{0} \leftrightarrow \ket{1}$ transition.  This spatial dependence is small if the microwave Rabi frequency $\Omega_R$ is fast compared to the optical depth $U_{optical} \approx \alpha_{eff} I_0 / c \epsilon_0 h$, but this is not a requirement - the heating depends only on the thorough mixing of the states $\ket{0}$ and $\ket{1}$

  Classical simulations predict a heating rate of $\sim$ 1.5 K s$^{-1}$ for the parameters shown in table \ref{parametertable} (figure \ref{classical_sim}).  As expected, simulations show this heating rate to be substantially robust to variations in trap parameters.  The robustness of the induced heating to details of the applied fields means multiple motional modes can be heated simultaneously; any motional mode in which the proper motion of the molecular ion is non-zero along the optical lattice axis will be heated.  



     It is the richness of the rotational state spectrum of polyatomic molecules that makes this type of `partial state determination' both feasible and valuable.  In a two state system, $\ket{0,1}$ spans the entire system, and the proposed measurement is worthless - it indicates nothing about which state within the $\ket{0,1}$ subspace we began in, and leaves the molecule in an unknown state within $\ket{0,1}$.  In contrast, when combined with a rich spectrum of accessible states, it yields both rapid measurement and heralded state preparation. Heralded \emph{single state} preparation requires repeated application of our method, as described in section \ref{singlestatesection}

\subsection{Why drive incoherently?}
 \label{whyincoherent}
The intentional scrambling of the state-dependent force appears counterintuitive.  Would it not be better to apply several carefully controlled  microwave drive  pulses, for example a series of $\pi$ pulses between $\ket{0}$ and $\ket{1}$?  In a world where the Hamiltonian could be controlled perfectly, $N$ such pulses could combine coherently, with the net momentum transferred scaling $\propto N$.   In the scheme proposed here, pulses only combine incoherently, and the net momentum transferred scales as $N^{1/2}$.  However, the freedom to ignore the detailed dynamics of the trapped ensemble allows the experimenter to exert much greater, and far less controlled, forces on the molecule, while simultaneously allowing the trap to `misbehave' essentially arbitrarily on the timescale of trap motion, $\tau \sim \omega_{trap}^{-1}$.  For example, the potentials shown in figure ~\ref{classical_sim}A are strongly nonlinear, but ensembles trapped in such potentials can still be heated by a force dithered on a timescale $O(\omega_{trap}^{-1})$. This freedom also allows the experiment to be driven in a non-perturbative limit, in contrast to the quantum-logic based proposals such as \cite{shi2013microwave}.

 $\Omega_{R}$ must be kept low enough that undesired transitions to other rotational states, for example  $\ket{0} \leftrightarrow \ket{2}$, are not driven.  These transitions are typically detuned by many GHz from $\omega_{01}$, easily allowing for $\Omega_{R} \gtrsim$ 10 MHz. 
High electrical Rabi frequencies $\Omega_{R}$ are technically straightforward - for example, 300 mV applied to electrodes 300 $\mu$m apart would realize electric-dipole allowed rotational transition rates with $\Omega_{R} \approx$ 10 MHz. 
Applied AC electric fields, either from the trap RF voltage or from the microwave drive, must be kept low enough that they do not substantially align molecules in states \emph{outside} of $\ket{0,1}$.  Molecules outside of $\ket{0,1}$ which are aligned while within the optical lattice will feel a time-dependent force, which will result in heating.  For the trap parameters listed in table \ref{parametertable}, this force is calculated to be about $10^3$ times lower than the time-dependent force responsible for the state-dependent heating for molecules within $\ket{0,1}$.  This parasitic heating is dominated by ``near resonant'' alignment of additional states by the microwave drive, with an effective detuning of $\sim$ 1 GHz, rather than low frequency trap fields.

 \subsection{Temperature readout}
 After the state-selective heating process described above, the temperature of the ensemble will be strongly dependent on the internal state of the molecular ion.  An ensemble temperature of a few mK will be reached in a few msec.  A final temperature of $\sim$ 2 mK would represent an occupation number of $n \approx 40$ for each of the motional modes of the crystal, and spatial extent of the Sr$^+$ ion of several microns.  Detecting this heating is therefore much easier than detecting the single motional quanta excited in QLS.  The temperature could be read out via Doppler thermometry on the narrow $^2$S$_{1/2}$ - $^2$D$_{5/2}$ transition; alternatively, the spatial motion of the Sr$^+$ ion could be resolved via the time-dependence of fluorescence from the ion illuminated with counter-propagating 422 nm $^2$S$_{1/2}$ - $^2$P$_{1/2}$ light beams.  Since the ion's motional extent is larger or comparable to the node spacing (211 nm), this fluorescence will be modulated at the trap secular 
frequency and higher sidebands\cite{Raabthermometry}.  The amplitude of these sidebands is highly temperature-dependent.

 \subsection{Extensions to larger subspaces}
 The simple scheme described above, in which $\ket{0}$ and $\ket{1}$ are assumed to be non-degenerate, is unrealistic for many molecules.  For example, if $\ket{0}$ and $\ket{1}$ are taken to be rotational levels,  there will be unresolved hyperfine levels within $\ket{0,1}$, from nuclear spins which couple only weakly (kHz or less) to the molecule's rotation.  Any such structure is far below the resolution of the rapidly driven microwave transitions used above. In addition, the scheme as described above imposes unreasonable constraints on the control of the polarization of applied electric and optical fields.  For example, if $\ket{1}$ is chosen to be $\ket{1_{010}}$, it is likely that population will leak into the states $\ket{1_{01\pm1}}$, even if all fields are nominally $\hat{z}$ polarized.

These limitations can be avoided simply by including the additional nearly-degenerate states in $\ket{0,1}$. Minimal modifications are required to extend the method to include additional states, replacing $\ket{0,1}$ with $\ket{i_1,i_2,..i_n}$.  The microwave fields that mix $\ket{0}$ and $\ket{1}$ must be supplemented by additional microwave fields that efficiently mix all states within $\ket{i_1,i_2,..i_n}$, and thus move the molecules between the effective potentials. For example states $\ket{1_{01\pm1}}$ can be mixed in with $\ket{0}$ and $\ket{1}$ via additional $\sigma^+$ or $\sigma^-$ microwave fields. The measurement leaves molecules that began within the subspace $\ket{i_1,i_2,..i_n}$  within this subspace, and the ensemble is heated; molecules in a state $\ket{k}$ outside $\ket{i_1,i_2,..i_n}$ remain in $\ket{k}$, and the ensemble is not heated. The heating rate depends only on the rate of flipping between states with different effective polarizability $\alpha_{eff}$, and will therefore be largely dependent of the number of states in $\ket{i_1,i_2,..i_n}$.  

Efficient projective measurement of the molecule's internal state from an initially thermally prepared molecule can be realized by intentionally including about half of the possible thermally occupied states in $\ket{i_1,i_2,..i_n}$. In the event of a ``yes'' measurement the molecule is known to lie within $\ket{i_1,i_2,..i_n}$; in the event of a ``no'' measurement the molecule is known to lie outside $\ket{i_1,i_2,..i_n}$.  In this case, additional microwave pulses can be applied, mixing additional states into $\ket{i_1,i_2,..i_n}$; for example, a microwave pulse which swaps $\ket{k} \leftrightarrow \ket{0}$ for an additional state $\ket{k}$. This can be repeated until a ``yes'' measurement is found.  Efficient searches are straightforward to design, and require no additional hardware beyond agile microwave electronics to implement.  Such searches can provide heralded state preparation into a single state from $n$ thermally occupied states in time $\mathcal{O}(\log{}n)$
     
\label{largesubspacesection}
 \subsection{Extensions to single state preparation}

  It is naturally desirable to extend this method to provide heralded \emph{absolute} state identification.  The following  sequence provides heralded projective measurement into a single arbitrary state $\ket{B}$:
  
  \begin{enumerate}
  \setlength \itemsep{-5pt}
  \item{As above, determine the state to lie within a manifold $\ket{i_1,i_2,..i_n}$}
  
  \item{With lattice lasers off, carefully drive a microwave transition between \emph{one} state $\ket{A}$ in $\ket{i_1,i_2,..i_n}$ and a single state $\ket{B}$ outside $\ket{i_1,i_2,..i_n}$.  This transition requires high ($< 1 $ KHz) resolution, and must be driven slowly. }
  
   \item{Check again to see if the molecule lies within the $\ket{i_1,i_2,..i_n}$ manifold. If it does not, the molecule is known to be in state $\ket{B}$.  If it does, we can assume that it is re-randomized within $\ket{i_1,i_2,..i_n}$, and we can repeat steps 2 and 3 until the molecule is projected into state $\ket{B}$}

    \end{enumerate}
    
    \label{singlestatesection}
    
 \subsection{Extensions to non-polar species}
 
 The microwave-frequency electric fields used to drive transitions between $\ket{0}$ and $\ket{1}$ are of course ineffective when applied to non-polar molecular ions, such as NO$_2^+$.  However, states $\ket{i_1,i_2,..i_n}$  can be chosen such that they are connected via a rotational Raman transitions, for example between $\ket{0_{000}}$ and $\ket{2_{020}}$.   These transitions can be driven via a single IR laser that is amplitude-modulated at the appropriate Raman transition frequency $\omega_{01}$, which can be chosen to be well within the bandwidth of high speed electro-optic modulators.  This laser could be the lattice laser itself, or an additional laser. 
 \label{ramansection}
 
  \subsection{Spectroscopy in larger ensembles}
The analysis above has been confined to a small crystal, comprised of a single molecular ion and a single atomic ion. Multiple molecular ions could also be co-trapped with one or more Sr$^{+}$ ion, and the state-dependent heating sequence described in section \ref{bolometrysection} would in that case heat the ensemble if one or more of the molecular ions was in the addressed subspace $\ket{i_1,i_2,..i_n}$.  The experiment would not be able to determine which trapped molecule was responsible for the heating, precluding meaningful projective measurement, but spectroscopy on the ensemble could still be performed.  In the simplest case, the experimenter would simply learn that \emph{some} thermally populated state is driven at the drive frequency $\omega_{01}$ - the same information that is learned in any single frequency spectroscopy experiment.  \label{bigcrystalsection}

 \section{Apparatus design and molecule choice}
 Existing ion trapping technologies are well suited to the proposal described here; in fact, the apparatus demonstrated by Chou et al.\cite{chou2016preparation} already realizes both Raman transitions and associated sideband cooling in molecular ions, albeit within the hyperfine manifold of a single rotational state.  That apparatus, which realized secular frequencies $\omega_{secular}$ of $\sim 2\pi \times$ 5 MHz, and AC Stark shifts of $\sim 2\pi \times 200$ kHz, is a natural starting point for design of a system.  We propose here a modestly weaker pseudopotential, realized via a macroscopic linear Paul trap, combined with a significantly stronger optical field.  The proposed trap design parameters are summarized in table \ref{parametertable}.  

\begin{table}[H]

\centering

\label{my-label} 
\begin{tabular}{|l|l|l|}
\hline
\textbf{parameter} & \textbf{symbol} & \textbf{value} \\ \hline 
electrode spacing & $d$                                           & 300 $\mu$m      \\
molecule mass &  $m_{molecule}$                                           & 76 amu     \\
atomic ion mass &  $m_{atom}$                                           & 88 amu      \\
molecule dipole moment &  $D$                                           & 2 Debye      \\

Trap RF frequency    & $\omega_{RF}$                                            & $2\pi \times$10 MHz          \\
Ion trap secular frequency & $\omega_{secular}$                                    & $2\pi \times$ 1 MHz           \\

 
Optical power &  $P$                                               & 1 W / beam   \\
Optical wavelength & $\lambda$     & 1050 nm         \\
Beam waist radius &  $\omega_0$                                     & 10 $\mu$m    \\
Rayleigh length & $z_R = \pi \omega_0^2 / \lambda$ & 0.3 mm  \\

Ion trap-lattice offset &  $\Delta z$                                     & $\lambda / 8 = 125$ nm   \\
Peak optical intensity & $I_{0} = 2P/\pi \omega_0^2$ & $6 \times 10^9 $ W/m$^2$  \\
 



Assumed molecular polarizability & $\bar{\alpha} = (\alpha_{\parallel} + 2\alpha_{\perp}))/3$									& 2 $\times 10^{-39}$ C m$^2$V$^{-1}$ [2$ \times 10^{-29}$ m$^3$ ] \\
Assumed polarizability anisotropy & $s = (\alpha_{\parallel} - \alpha_{\perp}) / \bar{\alpha} $ & 0.5  \\

Optical potential depth & $U_{optical} = \alpha_{eff} I_0 / c \epsilon_0 h$            & 7 MHz   \\
Lattice secular frequency (C.O.M. mode) & $\omega_{lattice} = \left(\frac{16 \pi^2 U_{optical}}{\lambda^2 m}\right)^{1/2}$            & 1.6 MHz   \\
Microwave drive amplitude &  $V_{mw} $           & 300 mV   \\
Microwave Rabi frequency &  $\Omega_R \approx V_{mw}D/dh    $        & 10 MHz   \\
Microwave dither rate &  $\Gamma_{flip}      $     & 2 MHz   \\
Effective polarizability for $\ket{0} = \ket{0_{000}}$ & $\alpha_{eff}(\ket{0})$ & 2 $\times 10^{-39}$ C m$^2$ V$^{-1}$ \\
Effective polarizability for $\ket{1} = \ket{1_{010}}$ & $\alpha_{eff}(\ket{1})$ & 1.7 $\times 10^{-39}$ C m$^2$ V$^{-1}$ \\  

Heating rate for molecules in $\ket{0}$ &        & 1.5 K s$^{-1}$   \\
Heating rate for molecules in $\ket{1}$ &        & 1.5 K s$^{-1}$   \\
Heating rate for molecules in $\ket{2} = \ket{1_{100}}$ &        & $<$ 1 mK s$^{-1}$   \\ \hline

\end{tabular}
\caption{Proposed parameters for a versatile hybrid trap for manipulation of polyatomic molecules.}
\label{parametertable}
\end{table}
 
 Because the only molecule-specific experimental design lies in the choice of the microwave pulse sequences applied to the trap electrodes, switching between molecules will be comparatively straightforward.  Two classes of molecular ions suggest themselves: positive analogs of stable, closed-shell molecules, (for example, 1,2-propanediol$^+$), and protonated versions of these molecules (for example, H-1,2-propanediol$^+$).  The principal difference between the two classes is that the protonated versions are typically closed shell $^1\Sigma$ states, simplifying their spectroscopy but making cooling of hyperfine degrees of freedom more challenging; molecules like propanediol$^+$ are necessarily open shell.  This gives the experimenter an additional handle to address the molecules - the magnetic field - but at the cost of greater complexity in $\hat{H}_{int}$. An additional advantage of open shell molecules is that an applied magnetic field breaks the $m_J$ degeneracy of rotational states, which otherwise must be addressed via carefully controlled polarization of electric and/or optical fields.  This effect is small ($\sim$ 1.4 MHz/Gauss) compared to $U_{optical}$ even for open shell molecules, and will not interfere with the proposed state selective heating. Open shell molecules are therefore well suited to \emph{absolute} single state preparation, including $m_J$ states and nuclear hyperfine structure. Open shell molecules are also much easier to produce than protonated molecules; indeed, in many cases the \emph{only} thing that is known about a molecular ion species is that it can produced from neutrals via a hot wire or electron bombardment, as is done in mass spectrometers.  Protonated species can generally be produced via collisions between neutral molecules and $H_3^{+}$, which readily donates its extra proton to almost any closed shell neutral species\cite{burt1970some}
 

 
 \section{Applications}
 The heralded single state preparation described above forms the heart of a broad set of experiments, which can exploit the diverse physics enabled by polyatomic molecules.  The basic sequence of such an experiment is as follows:
 
 \begin{enumerate}
 \item The internal state of the molecular ion is measured as described above. 

  \item The state of the molecule is manipulated by applying microwave-frequency voltages directly to the trap electrodes. An arbitrary unitary transformation $\mathrm{U}$ among a chosen set of molecular states can be applied via appropriate choice of applied voltages. The optical lattice is turned off during this time.

  \item The internal state of the molecular ion is read out again.
  \end{enumerate}
 
  Depending on the choice of the unitary transformation $\mathrm{U}$, this sequence will realize high-precision spectroscopy of the molecular ion in question, non-destructive single-molecule identification and chiral readout, or a low-decoherence quantum information platform in which a single molecule comprises several Qubits. 
 
 \subsection{\textbf{High resolution spectroscopy}}
 If $\mathrm{U}$ is chosen to be a carefully applied exchange between the prepared state $\ket{A}$ and another state, the above sequence realizes a high-resolution spectrometer.  $\mathrm{U}$ could be a Rabi pulse or a Ramsey sequence.  Trapped molecular ions in a cryogenic environment are remarkably well isolated from the environment, and are expected to exhibit exceptionally long coherence times, allowing for high resolution spectroscopy.  Background gas collision rates far below 1 Hz are possible, and in closed-shell molecules all rotational transitions are Zeeman-insensitive. The dominant broadening mechanism is expected to be unwanted Stark shifts from the time-varying electric fields from the trap electrodes.  The ions by definition find locations where $\braket{\vec{E}} = 0$, but in general $\braket{E^2} > 0$, and unwanted Stark shifts are a possibility.  The DC polarizability of rotational states in asymmetric top molecules varies over many orders of magnitude, depending on the proximity of the nearest state of opposite parity, but in a well compensated trap and a small crystal, in general $\abs{E} < 1 $ V cm$^{-1}$, and Stark shifts for most states are expected to be on the 10 Hz level or below. Unwanted \emph{light shifts} from the lattice lasers are  not an issue, since these lasers are turned off during the spectroscopy pulse.
 

 \subsection{\textbf{Chiral readout}}

The experimental sequence described above allows for definitive identification of the chirality of the trapped molecular ion.  Such an experiment would proceed as follows: as before, the molecule is initialized into a known rotational state $\ket{A}$. As demonstrated in \cite{eibenberger2017enantiomer}, the molecule can then be transferred to a different state $\ket{B}$ conditional on its chirality. To do this, the molecule is transferred simultaneously via two distinct paths: directly from $\ket{A} \rightarrow \ket{B}$, and indirectly $\ket{A} \rightarrow \ket{C} \rightarrow \ket{B}$.   The relative phases of the paths can be chosen to interfere constructively for one enantiomer and destructively for the other; a change in the phase of any of the applied fields reverses the choice of enantiomer.  Readout of the state thus reads out the molecule's chirality.

It is notable that this enantiomer-selective measurement is non-destructive; the molecule is left in the trap, and in fact is left in a known state. The molecule can thus be re-measured milliseconds - or minutes - later.  Many neutral species, for example HOOH, HSSH, and analogous species, are known to tunnel between chiral conformers.  
The tunneling time can vary from nanoseconds to beyond the age of the universe, but has never been observed at rates slower than the resolution of beam-based microwave spectroscopy ($\sim$ 5 kHz).  The experiment proposed here could push this by several orders of magnitude, allowing both an exquisite probe of intra-molecular dynamics, and an ideal platform to look for predicted but never before seen energy differences between enantiomers arising from the short-range nuclear weak force \cite{quack2012molecular}.


\subsection{\textbf{Searches for new physics}}

Ultraprecise spectroscopy of diatomic molecules has emerged as a sensitive probe of physics beyond the Standard Model.  A dramatic example is the search for a permanent electron electric dipole moment (eEDM); while the eEDM is predicted to be unobservably small in the standard model, many plausible extensions predict a larger value, and current experimental limits strongly limit such extensions.  The high internal electric field in molecules makes them exquisitely sensitive to the eEDM, and the current limit on the eEDM was set recently in precision spectroscopy experiments on neutral ThO molecules\cite{baron2014order}; a comparable limit was recently established using HfF$^+$ ions \cite{PhysRevLett119153001}. Polyatomic molecules have recently been proposed as leading candidates in next-generation searches for both eEDMs and related nuclear EDMs \cite{kozyryev2017precision}. The generality of the method proposed here is a significant advantage in these schemes, which to date rely on a favorable electronic structure in the molecules which allows for laser induced fluorescence or photofragmentation.


Molecular ions have also been proposed as candidates for searches for variation in the electron/proton mass ratio.  In such an experiment, a high-precision clock based on nucleon mass - for example, on a vibrational transition - would be compared to a more conventional optical clock based on electron mass.  The homonuclear O$_2^+$ ion has been suggested as a candidate in such searches\cite{hanneke2016high}.  Raman-active modes of nonpolar, closed shell triatomic molecular ions such as NO$_2^+$, which could be observed directly by the system described in this work, appear to be a highly systematic-immune platform for such searches.  Since these molecules are non-polar, this experiment would require the Raman-mediated heating described in section \ref{ramansection}.  

\subsection{\textbf{Quantum information}}

The rich, controllable states of molecules suggests their potential as a quantum information platform.  In such a scheme, ``arbitrary single-qubit rotations'' of traditional 2-state qubits would be replaced by ``arbitrary unitary transformations within a subset of low-lying rotational states''.  The operations that can be applied to such a molecule are a strict superset of NMR rotations; this platform could be thought of as a revisit of NMR quantum computing, with \emph{additional degrees of freedom} (rotation), \emph{additional controls} (microwave frequency electric fields), and \emph{definitive initialization}.  Can meaningful error correction be applied within single molecules?  Can a single molecule be transported, carrying multiple qubits?   While answering such questions is beyond the scope of the research proposed here, they illustrate the potential power of molecular ions as low-decoherence quantum machines.  

  Scalability is of course a major concern in any quantum information proposal.  Here, distinct, co-trapped molecules - either identical or distinct - could be entangled via shared phonon modes, as in traditional atomic ion quantum information schemes.  Such a scheme requires substantially more experimental control than the incoherent heating required in this proposal; in particular, well controlled, sideband resolved manipulation and ground state cooling are required. 
  
 \section{Complexity and challenges}
\begin{figure}[h!]
\includegraphics[width=6in]{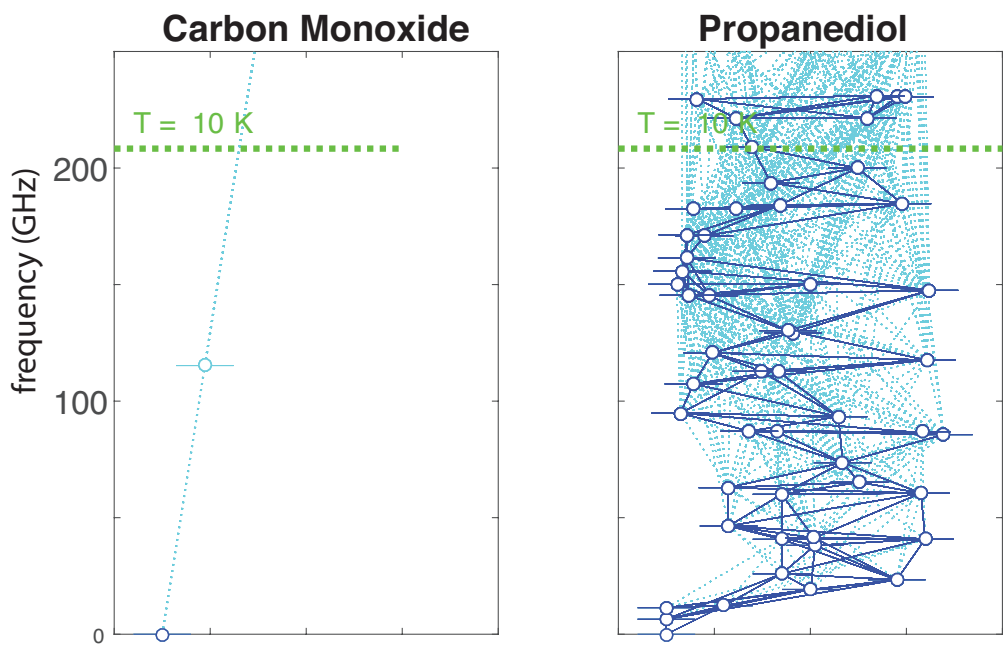}
\caption{The low-lying rotational states of carbon monoxide (left) and 1,2-propanediol (right).  Electric dipole allowed transitions below 20 GHz are marked in dark blue.  Allowed transitions above 20 GHz are marked in light blue. All thermally occupied states in the right hand molecule can be reached via combinations of low frequency, electric-dipole allowed transitions.
\label{biggrot}}
\end{figure}

Polyatomic molecules are undoubtedly complex.  Figure \ref{biggrot} compares the rotational Grotrian diagram for carbon monoxide and 1,2-propanediol up to 10 Kelvin.  It might appear at first that  controlling the intricacy of the more complex molecule is impossible, but in fact it is \emph{easier} to manipulate the state of propanediol than carbon monoxide.  This is because agile arbitrary waveform generators and microwave electronics, with bandwidth up to $\sim$40 GHz, can address every level populated in a 10 Kelvin propanediol molecule - in contrast,  carbon monoxide, and in fact almost all diatomic molecules, require challenging electronics operating at frequencies above 100 GHz. Several groups propose addressing this challenge directly, via comb-mediated coherent control of laser fields differing by THz \cite{leibfried2012quantum,ding2012quantum,chou2016preparation}. In contrast, the only lasers required for the larger molecules proposed here are a single, non-tunable infrared laser, and the cooling lasers for the co-trapped Sr$^+$. 

While the diagram in figure \ref{biggrot}B appears incomprehensible, in fact the spectroscopy of such molecules is straightforward; all the lines in neutral 1,2-propanediol shown in this figure have been observed and unambiguously assigned.  Spectroscopy of polyatomic ions is far less developed than spectroscopy of neutral species; while the form of the rotational Hamiltonian is unchanged from neutral molecules, the exact rotational constants and internal coupling constants will need to be measured.  High-sensitivity rotational spectroscopy via buffer gas cooling has proven a versatile technique on neutral molecules, and could be extended to cold plasmas of untrapped ionic species\cite{mefirstFTMW}.   In addition, to our knowledge the polarizability tensor of all polyatomic molecular ions is unknown, and will require measurement; these measurements can of course be performed using the same apparatus that is described here.

  While the Hamiltonian of a rigid molecule in arbitrary electric, magnetic, and optical fields can be written down exactly, the complexity of the Hamiltonian quickly grows beyond direct human understanding.  We therefore rely on numerical simulation.  Our home built software (MATLAB) uses the PGOPHER package to calculate $H_{internal}$, electric dipole transition matrices, and (for open shell molecules) magnetic dipole transition matrices\cite{PGOPHER}.  Optically driven Raman transitions and the motional Hamiltonian are calculated directly.  Arbitrary pulse sequences of electric, magnetic, and optical fields, either uniform or spatially varying, can be simulated.  Various choices of applied pulses yield simulations of standard microwave free induction decay spectroscopy, chirally sensitive three wave mixing, trap motional dynamics, sideband cooling, or state-specific detection.  
  
   The complex dynamics of a warm ($T \gg$ 1 mK) ensemble trapped in the potentials shown in figure \ref{classical_sim} would require prohibitive computing power for a fully quantum simulation.  We therefore calculate heating rates via a semi-classical approximation, where internal molecular dynamics are simulated under the assumption of spatially uniform fields, and motional dynamics are simulated classically.  Figure \ref{classical_sim} shows the results of one such semi-classical simulation. 
  
  \section{Conclusion}
  Polyatomic molecular ions contain great potential as highly controllable, low-decoherence quantum systems.  The tools presented here will allow preparation, detection, and characterization of molecular ions, including high-fidelity non-destructive readout and heralded state preparation at the single quantum state level. These tools are the prerequisite for a broad suite of experiments in high precision spectroscopy and quantum control.
  
 \clearpage
 \bibliographystyle{ieeetr}
\bibliography{QC.bib}

\end{document}